\documentclass{article}
\usepackage{spconf,amsmath,graphicx,hyperref}
\usepackage{xeCJK}
\usepackage{amssymb}
\usepackage{booktabs}
\usepackage{multirow}
\usepackage{makecell}

\title{XINGCHEN-UNIFE: A UNIFIED GENERATIVE SPEECH FRONT-END VIA COMPLEMENTARY ACOUSTIC-SEMANTIC MODELING}

\name{Kelan Kuang \sthanks{E-mail: kuangkl@chinatelecom.cn}, Qihan Hu, Jian Kang, Jie Li, Yongxiang Li,
Yi Liu\sthanks{Corresponding author: Yi Liu (e-mail: liuyi@chinatelecom.cn).}}

\address{China Telecom Artificial Intelligence Technology (Beijing) Co., Ltd.}
\makeatletter
\newcommand{\bstctlcite}[1]{%
  \@bsphack
  \if@filesw
    \immediate\write\@auxout{\string\citation{#1}}%
  \fi
  \@esphack
}
\makeatother

\begin{document}
\bstctlcite{IEEEcontrol}
%
\maketitle
\linespread{0.99}\selectfont
\begin{abstract}

This paper presents XingChen-UniFE, a unified large audio language model (LALM) for general speech restoration (GSR) and target speaker extraction (TSE). 
The proposed model employs complementary semantic and acoustic representations through a dual-stream encoder, where a self-supervised learning model provides hierarchical semantic features and a learnable acoustic branch captures fine-grained structures from complex spectra. 
A bidirectional cross-attention module further facilitates information exchange between the two representations. 
The fused representation conditions the LALM for discrete speech token generation with task and speaker information. 
Experiments demonstrate strong performance on both GSR and TSE and robust generalization to unseen degradation conditions.

\end{abstract}

\section{INTRODUCTION}

Real-world speech is often degraded by noise, reverberation, transmission distortion, and interfering speakers, which reduces speech quality and intelligibility. Existing systems typically use separate models for different speech restoration tasks and target speaker extraction, increasing training and deployment costs. This motivates general speech restoration with a single model.

Most general speech restoration (GSR) methods follow discriminative paradigms by predicting waveforms, spectrograms, or time-frequency masks \cite{zheng2023sixty,wang2018supervised,ochieng2023deep}. Although effective for specific tasks, they usually rely on task-specific architectures and objectives, making heterogeneous restoration tasks difficult to handle within a unified framework. Severe degradation further requires modeling acoustic details, linguistic content, and speaker characteristics beyond signal reconstruction alone.

Recent advances in large audio language models (LALMs) and neural audio codecs enable speech processing to be formulated as conditional discrete token generation \cite{borsos2023audiolm}. However, LALM-based restoration requires input representations that preserve robust semantic information and fine-grained acoustic cues. Self-supervised learning (SSL) models, such as wav2vec 2.0 \cite{baevski2020wav2vec}, HuBERT \cite{hsu2021hubert}, and WavLM \cite{Chen2022WavLM}, provide hierarchical speech representations but are not optimized for waveform reconstruction and may lose fine acoustic details. Recent SR studies therefore explore joint acoustic-semantic modeling to reduce linguistic and acoustic hallucinations. PASE \cite{rong2026pase} projects low-level SSL representations into the space of high-level phonetic representations and combines them by addition, while L3-SE \cite{wang2026reducing} jointly distills acoustic and semantic representations from WavLM and uses them as continuous LM conditions. 
However, their acoustic representations remain largely derived from SSL features and thus lack explicit modeling of the frequency and phase structure of degraded speech. Complex spectrogram, by contrast, retains both a physically meaningful frequency axis and phase information, providing a natural representation for frequency-aware acoustic modeling. Moreover, existing acoustic and semantic representations are typically fused by simple addition or concatenation, without explicit interaction.

In this work, we propose XingChen-UniFE, a unified LALM-based model for GSR and TSE that integrates complementary acoustic and semantic representations. The proposed model employs a dual-stream encoder consisting of an SSL-based semantic branch and an acoustic branch, followed by a bidirectional cross-attention module to explicitly model the interaction between the two representations. The resulting fused representation provides the LALM with both robust linguistic cues and fine-grained acoustic details for discrete speech token generation. With task and speaker conditions, XingChen-UniFE can handle diverse speech restoration and target speaker extraction tasks in a single model.

\begin{figure}[htb]
    \centering
    \includegraphics[width=0.49\textwidth]{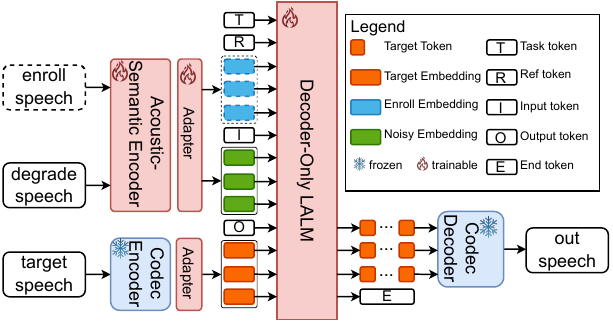}
    \caption{Overall architecture of XingChen-UniFE. Enrollment speech is used only for TSE.}
    \label{fig:overview}
\end{figure}

The contributions of this work are summarized as follows:
\textbf{1)} We propose XingChen-UniFE, a unified LALM-based model that supports speech restoration and TSE under diverse degradation conditions.
\textbf{2)} We design a dual-stream acoustic-semantic encoder that combines SSL representations with complex spectral features to capture complementary linguistic and acoustic information.
\textbf{3)} We introduce a bidirectional cross-attention module to explicitly model interactions between acoustic and semantic representations.

\section{Proposed Method}

\subsection{Overall Architecture}
XingChen-UniFE formulates GSR and TSE as conditional discrete speech generation. Given degraded speech \(x\), a task control token, and enrollment speech \(r\) for TSE, the model predicts acoustic and semantic tokens of the target speech.

As shown in Fig.~\ref{fig:overview}, the model comprises a complementary acoustic-semantic encoder, an LALM backbone, and a neural audio codec. The encoder produces the conditional representation, which guides the LALM to generate interleaved delayed tokens in an autoregressive manner. The generated tokens are then rearranged and decoded by the codec decoder.

\subsection{Complementary Representation Learning}
\begin{figure}[htb]
    \centering
    \begin{minipage}{\linewidth}
        \centering
        \includegraphics[width=\linewidth]{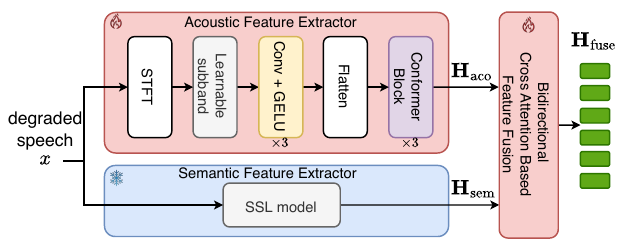}
        \centerline{(a)}
    \end{minipage}

    \vspace{0.3em}

    \begin{minipage}{\linewidth}
        \centering
        \includegraphics[width=0.85\linewidth]{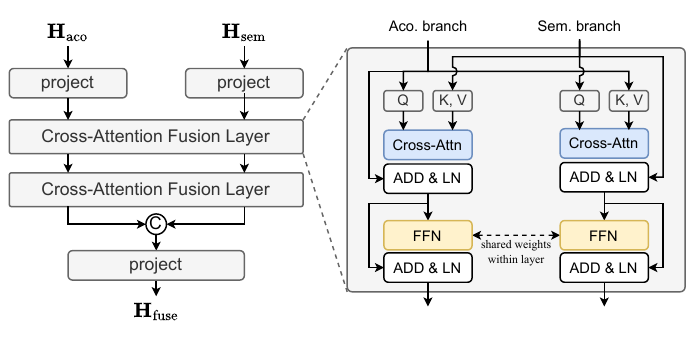}
        \centerline{(b)}
    \end{minipage}

    \caption{Overview of the complementary acoustic-semantic encoder.
    (a) Semantic and acoustic feature encoding;
    (b) bidirectional feature fusion.}
    \label{fig:aco_sem_encoding}
\end{figure}

As shown in Fig.~\ref{fig:aco_sem_encoding}, the complementary encoder consists of parallel semantic and acoustic branches followed by a bidirectional feature fusion module.

\subsubsection{Semantic and Acoustic Feature Encoding}

The semantic branch employs a pre-trained SSL model to extract speech representations. Given the input waveform $x$, we average the hidden representations from all layers of the SSL model:
\begin{equation}
\mathbf H_{\mathrm{sem}}
=
\frac{1}{N_w}
\sum_{\ell=1}^{N_w}
\operatorname{SSL}_{\ell}(x),
\label{eq:semantic_representation}
\end{equation}
where $N_w$ denotes the number of Transformer layers in the SSL model. 
To complement SSL features with fine-grained acoustic details, we introduce a learnable acoustic branch. The input waveform is transformed into a complex STFT spectrogram \(\mathbf X\in\mathbb C^{T\times321}\), whose real and imaginary components are stacked and mapped through a sub-band projection. The resulting features are processed by three 2D convolutional blocks and three Conformer \cite{gulati2020conformer} blocks to obtain the acoustic representation \(\mathbf H_{\mathrm{aco}}\). The convolutional blocks reduce the frequency dimension while preserving temporal resolution.

\subsubsection{Bidirectional Feature Fusion}

To facilitate information exchange between the acoustic and semantic representations, we propose a two-layer bidirectional cross-attention fusion module. As shown in Fig.~\ref{fig:aco_sem_encoding}(b), the semantic representation $\mathbf H_{\mathrm{sem}}$ and acoustic representation $\mathbf H_{\mathrm{aco}}$ are first mapped into a shared embedding space through separate linear projections.

In each fusion layer, one branch serves as Query and the other as Key and Value, with the roles exchanged for bidirectional cross-attention. Each output is followed by a residual connection and Layer Normalization, and then by a shared FFN with another residual connection and Layer Normalization. After two fusion layers, the two outputs are concatenated and linearly projected to obtain \(\mathbf H_{\mathrm{fuse}}\).

\begin{table*}[t]
    \caption{Performance on Libri2Mix for Denoise, TSE, and TSE\&Denoise using the \textit{mix-single}, \textit{mix-clean}, and \textit{mix-both} subsets, respectively. Best and second-best results are \textbf{bold} and \underline{underlined}.}
    \centering
    \begin{tabular}{lcccc cccc cccc}
        \toprule
        
        & \multicolumn{4}{c}{Denoise}
        & \multicolumn{4}{c}{TSE} 
        & \multicolumn{4}{c}{TSE\&Denoise} \\
        \cmidrule(lr){2-5}
        \cmidrule(lr){6-9}
        \cmidrule(lr){10-13}
        & PESQ & STOI & SBS & LPS
        & PESQ & STOI & SBS & LPS
        & PESQ & STOI & SBS & LPS \\
        \midrule

        LLaSE-G1
        & 1.218 & 74.23 & .835 & .823
        & 1.127 & 66.81 & .774 & .654 
        & 1.091 & 61.04 & .708 & .456 \\

        UniSE
        & 1.683 & 87.47 & .884 & .869
        & 1.666 & \textbf{84.18} & .882 & \textbf{.814} 
        & 1.377 & 76.63 & .815 & .620  \\

        UniTok
        & \underline{2.100} & \underline{90.14} & \underline{.904} & \underline{.894}
        & \underline{2.015} & \underline{83.71} & \underline{.887} & \underline{.797} 
        & 1.640 & \textbf{79.22} & \underline{.845} & \underline{.678} \\

        \textbf{Proposed}
        & \textbf{2.137} & \textbf{90.56} & \textbf{.916} & \textbf{.930}
        & \textbf{2.093} & 81.74 & \textbf{.888} & .786 
        & \textbf{1.703} & \underline{78.55} & \textbf{.852} & \textbf{.714} \\

        \bottomrule
    \end{tabular}
    \label{tab:libri2mix_sr_tse}
\end{table*}

\subsection{Conditional Token Generation}

During training, the target speech \(s\) is encoded by a frozen Codec encoder into acoustic tokens \(\mathbf A=\{A_{t,k}\}\) and semantic tokens \(\mathbf S=\{S_{t,k}\}\), where \(t\) is the temporal index and \(k=0,\ldots,K-1\) is the codebook index.

To jointly model the two token types, acoustic and semantic tokens at the same temporal position are first interleaved:
\begin{equation}
C_{2t,k}=A_{t,k},
\qquad
C_{2t+1,k}=S_{t,k}.
\end{equation}
The interleaved sequence is then rearranged using the delayed pattern~\cite{copet2023simple} to form the autoregressive target sequence:
\begin{equation}
\widetilde{C}_{u+k,k}=C_{u,k},
\qquad
k=0,\ldots,K-1 ,
\label{eq:delayed_pattern}
\end{equation}
where $u$ denotes the position in the interleaved sequence.

The fused representation $\mathbf H_{\mathrm{fuse}}$ is mapped to the LALM input space through a linear adapter to obtain the conditional representation $\mathbf H_X$. For TSE, the enrollment speech $r$ is processed by the same encoder and projection pipeline to obtain $\mathbf H_R$. Conditioned on $\mathbf H_X$, $\mathbf H_R$, the task control token $\mathcal S$, and previously generated tokens, the LALM backbone performs conditional next-token prediction,
\begin{equation}
\mathbf H_u
=
f_\theta
\left(
\widetilde{\mathbf C}_{<u},
\mathbf H_X,
\mathbf H_R,
\mathcal S
\right),
\label{eq:lalm_hidden_state}
\end{equation}
where $\widetilde{\mathbf C}_{<u}$ denotes the tokens preceding generation step $u$. Based on $\mathbf H_u$, $K$ parallel output heads predict the token distributions over the corresponding codebooks, with greedy decoding used during inference. The generated sequence is rearranged into time-aligned acoustic and semantic tokens and decoded by the codec decoder.

\section{EXPERIMENTAL SETUP}
\label{sec:experimental_setup}
\subsection{Implementation Details}

The proposed XingChen-UniFE adopts a decoder-only Transformer as the LALM backbone, which comprises 28 layers with a hidden dimension of 1024 and grouped-query attention \cite{ainslie2023gqa} with 16 query heads and 8 key-value heads. The dual-stream encoder consists of a frozen semantic branch based on WavLM-Base-Plus \cite{Chen2022WavLM} and a trainable acoustic branch, while H-Codec \cite{liu2025quarkaudiotechnicalreport} serves as the speech codec. The LALM backbone is trained from scratch, while WavLM and H-Codec remain frozen, resulting in approximately 502 M trainable parameters. The model is optimized for 400k steps with a batch size of 64 using AdamW optimizer \cite{loshchilov2019decoupled} and a Noam schedule with a 4k-step warm-up \cite{vaswani2017attention}.

\subsection{Training Data}

Degraded speech is simulated on the fly. Each sample is randomly corrupted by one or more degradations, including additive noise, reverberation, interfering speech, clipping, packet loss, and bandwidth limitation. Target speech is sampled from DNS Challenge \cite{dubey2024icassp},
Emilia \cite{he2024emilia}, LibriSpeech \cite{panayotov2015librispeech}, LibriHeavy \cite{kang2024libriheavy}, and AISHELL-3 \cite{shi2020aishell}. Interfering speech is collected from WenetSpeech4TTS \cite{ma2024wenetspeech4tts}, VoxCeleb2 \cite{chung2018voxceleb2}, and CN-Celeb \cite{fan2020cn}, while noise samples are obtained from DEMAND \cite{thiemann2013diverse}, WHAM! \cite{wichern2019wham}, DNS Challenge, RealMAN \cite{yang2024realman}, ESC-50 \cite{piczak2015esc}, and DESED \cite{turpault2019sound}. RIRs are randomly sampled from SLR26 and SLR28 \cite{SLR26}. Utterances from Emilia and LibriHeavy with DNSMOS scores below 3.5 are discarded. 
The resulting training pool comprises approximately 10k hours of target speech, 405k hours of noise, and 4640 hours of interfering speech. All audio samples are resampled to 16 kHz and segmented into clips ranging from 3 to 15 seconds in duration.

\begin{table*}[t]
    \caption{GSR subtask performance measured by LPS$\uparrow$ / WER (\%)$\downarrow$. Mixed denotes samples containing all listed distortion types.}
    \label{tab:sr_subtasks}
    \centering
    \small
    \setlength{\tabcolsep}{4.5pt}
    \begin{tabular}{lcccccccc}
        \toprule
        Model & Noise & Rev. & Interf. & BW & Clip & Loss & Mixed & Avg. \\
        \midrule

        UniSE
        & .754/21.8 & .833/11.1 & .884/7.1 & .846/10.2
        & .873/6.4 & .814/13.4 & .335/89.1 & .763/22.7 \\

        UniTok
        & \underline{.805}/\underline{15.4} & \underline{.836}/\underline{6.7} & \underline{.897}/\underline{5.7} & \underline{.853}/\underline{7.8}
        & \underline{.892}/\underline{4.3} & \underline{.828}/\underline{10.3} & \underline{.472}/\underline{68.5} & \underline{.798}/\underline{17.0} \\

        \textbf{Proposed}
        & \textbf{.840}/\textbf{10.8}
        & \textbf{.859}/\textbf{5.1}
        & \textbf{.903}/\textbf{5.2}
        & \textbf{.881}/\textbf{4.8}
        & \textbf{.899}/\textbf{3.5}
        & \textbf{.852}/\textbf{8.8}
        & \textbf{.548}/\textbf{55.4}
        & \textbf{.826}/\textbf{13.4} \\
        \bottomrule
    \end{tabular}
\end{table*}

\begin{figure*}[t]
    \centering
    \includegraphics[width=0.99\textwidth]
    {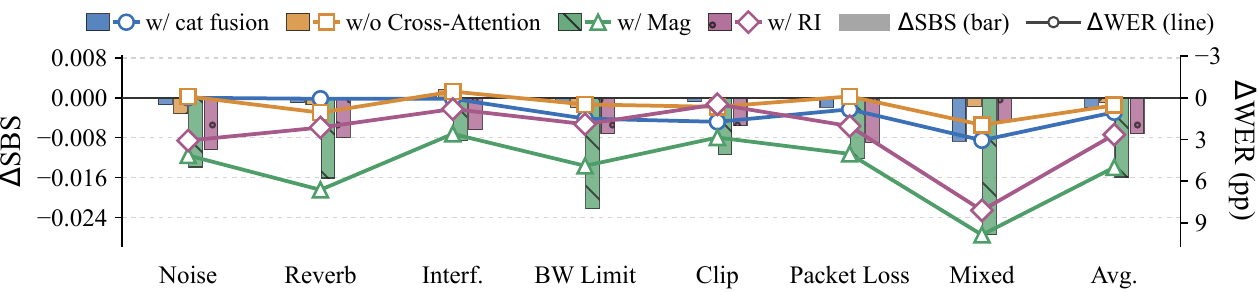}
    \caption{Task-wise ablation of the fusion and acoustic representations. Bars and lines denote $\Delta$SBS and $\Delta$WER
    relative to XingChen-UniFE, respectively; downward displacement indicates
    performance degradation.}
    \label{fig:taskwise_ablation}
\end{figure*}



\section{RESULTS}

We evaluate XingChen-UniFE and the baselines on GSR and TSE using PESQ \cite{rix2001perceptual}, STOI \cite{taal2011algorithm}, SpeechBERTScore (SBS) \cite{saeki24_interspeech}, WER, and Levenshtein Phoneme Similarity (LPS) \cite{Pirklbauer+23}. Higher values are better except for WER.

Additional evaluation results are provided in Appendix~\ref{app:additional_results}. To facilitate reproducibility, we make the model weights, evaluation datasets, and scripts publicly available.

\subsection{Comparison Experiment}
XingChen-UniFE is compared with three generative multi-task baselines: LLaSE-G1 \cite{kang2025llase}, UniSE \cite{yan2025unise} and UniTok \cite{liu2025unitok} for GSR and TSE. LLaSE-G1 and UniSE use the publicly released checkpoints provided by the original authors, whereas UniTok is reproduced by us and trained under the same conditions as the proposed model.

Table~\ref{tab:libri2mix_sr_tse} compares the models on Libri2Mix. XingChen-UniFE achieves the best results on all metrics for the denoising task, and leads in PESQ and SBS for TSE. Under the more challenging TSE\&Denoise condition, it achieves the best PESQ, SBS, and LPS, demonstrating strong performance across different restoration tasks.

Table~\ref{tab:sr_subtasks} further compares individual GSR subtasks. XingChen-UniFE consistently achieves the best LPS and WER, with a larger margin under the Mixed condition, indicating improved robustness to multiple degradations.

We further evaluate generalization to two unseen degradation types excluded from training, codec distortion and wind noise, as shown in Table~\ref{tab:unseen}. XingChen-UniFE achieves the best SBS and WER for codec distortion and the best LPS, SBS, and WER for wind noise, demonstrating strong generalization to unseen conditions.

\begin{table}[t]
    \caption{Performance under unseen degradation.}
    \label{tab:unseen}
    \centering
    \small
    \setlength{\tabcolsep}{5pt}
    \begin{tabular}{lccc ccc}
        \toprule
        &\multicolumn{3}{c}{Codec}
        &\multicolumn{3}{c}{Wind Noise}\\
        \cmidrule(lr){2-4}
        \cmidrule(lr){5-7}
        &LPS&SBS&WER
        &LPS&SBS&WER\\
        \midrule

        LLaSE-G1
        &.824&.855&8.70
        &.834&.852&8.16\\

        UniSE
        &\textbf{.900}&.909&5.45
        &\underline{.896}&.903&3.91\\

        UniTok
        &.896&\underline{.937}&\underline{3.29}
        &.880&\underline{.920}&\underline{3.80}\\

        \textbf{Proposed}
        &\underline{.896}&\textbf{.938}&\textbf{3.25}
        &\textbf{.900}&\textbf{.927}&\textbf{3.20}\\

        \bottomrule
    \end{tabular}
\end{table}

\subsection{Ablation Study}

\begin{table}[t]
    \caption{Ablation of the acoustic and semantic branches on Libri2Mix \textit{mix-single}.}
    \centering
    \small
    \begin{tabular}{lccccc}
        \toprule
        & PESQ & STOI & SBS & LPS & WER \\
        \midrule
        \textbf{Proposed} & \textbf{2.14} &\textbf{90.6} &\textbf{.916} &\textbf{.930} &\textbf{7.74} \\
        \quad w/o Sem. &\underline{2.09} & \underline{90.2} &.902 &.892 & 13.01\\
        \quad w/o Aco. &1.84 & 89.1 &\underline{.911} &\underline{.917} & \underline{9.03}\\
        
        \bottomrule
    \end{tabular}
    \label{tab:taskwise_ablation}
\end{table}

Table~\ref{tab:taskwise_ablation} examines the contributions of the two branches. Removing the semantic branch causes larger degradation in SBS, LPS, and WER, whereas removing the acoustic branch mainly affects PESQ and STOI. As the former metrics emphasize linguistic consistency and the latter reflect perceptual quality and intelligibility, these results demonstrate the complementary roles of the semantic and acoustic branches.

Fig.~\ref{fig:taskwise_ablation} shows task-wise ablations of feature fusion and acoustic representation. Both simple concatenation (cat Fusion) and removal of cross-attention (w/o C.A.) degrade performance, validating explicit bidirectional interaction. Moreover, the acoustic-only variant using real and imaginary components (w/ RI) clearly outperforms its magnitude-only counterpart (w/ Mag), highlighting the importance of phase information, particularly under mixed degradations.

\section{CONCLUSION}
This paper presents XingChen-UniFE, a discrete speech generation model for GSR and TSE. The model integrates complementary acoustic and semantic representations through a dual-stream encoder and bidirectional cross-attention. This design jointly models fine-grained acoustic structure and semantic information. The results demonstrate the effectiveness of the proposed model and further validate the proposed acoustic-semantic modeling and fusion strategy.

\clearpage



\bibliographystyle{IEEEtran}
\bibliography{refs}

\clearpage
\onecolumn
\linespread{1}\selectfont
\raggedbottom
\appendix
\section{Additional Evaluation Results}
\label{app:additional_results}

\setcounter{table}{0}
\renewcommand{\thetable}{A.\arabic{table}}
\renewcommand{\theHtable}{appendix.\arabic{table}}

\makeatletter
\newenvironment{appendixtable}{%
  \par\addvspace{10pt}\noindent
  \begin{minipage}{\linewidth}
  \def\@captype{table}%
  \centering\small
  \setlength{\tabcolsep}{4pt}%
  \renewcommand{\arraystretch}{1.15}%
}{%
  \end{minipage}\par\addvspace{6pt}%
}
\makeatother

\subsection{Evaluation Setup}

\textbf{UniFE model weights.} All UniFE results reported in this appendix are obtained using a different set of model weights from those used in the main text. The UniFE model evaluated here is trained on a curated set of high-quality speech data.

We evaluate UniFE and the following comparison systems on Libri2Mix, EARS-Reverb, EARS-WHAM, WHAM!, WHAMR!, and PLC-Challenge: UniSE, an open-source generative speech restoration model\footnote{\url{https://github.com/alibaba/unified-audio/tree/main/QuarkAudio-UniSE}}; AuK, a speech generation and editing model\footnote{\url{https://github.com/Tencent-Hunyuan/AuK}}; LauraTSE, a target speaker extraction model\footnote{\url{https://github.com/Beilong-Tang/lauraTSE_code}}; SAM-Audio, a speech separation model\footnote{\url{https://github.com/facebookresearch/sam-audio}}; and Auphonic, a commercial AI audio enhancement system.

In addition to these test sets, we construct a MultiDistortion (MD) test set using clean speech from high-quality recordings in publicly available speech datasets. This test set allows us to compare the ability of different models to restore the same speech content under different distortions. We select 25 utterances each from AIShell3, ABCS AC, EARS, VCTK, and Libri2Mix S1, yielding 125 utterances covering Chinese and English. All recordings are mono audio sampled at 16~kHz, with durations of 8--12~s. For each clean utterance, we create six individual distortion conditions---additive noise, reverberation, interfering speech, bandwidth limitation, clipping, and packet loss---as well as a combined condition containing all six distortions. Each of the seven conditions contains 125 input--reference pairs, resulting in 875 pairs in total. Additive noise is drawn from DEMAND, WHAM!, and NoiseX92. Reverberation is simulated using synthetic room impulse responses (RIRs) from OpenSLR SLR26 and SLR28, as well as measured RIRs from ACE-Challenge. In the combined condition, interfering speech, reverberation, and noise are introduced sequentially, followed by bandwidth limitation, clipping, and packet loss. For the reverberation and combined conditions, the reference is obtained by convolving the target speech with the early RIR, truncated at 5~ms after the peak with the largest absolute amplitude. Clean speech serves as the reference for the remaining conditions.

Tables~\ref{tab:app_01}--\ref{tab:app_05} report results for speech denoising and joint denoising and dereverberation; Tables~\ref{tab:app_06}--\ref{tab:app_11} report TSE results; Tables~\ref{tab:app_12}--\ref{tab:app_20} report results for the remaining individual distortion restoration tasks; and Table~\ref{tab:app_21} reports results for restoration under combined distortions. The evaluation metrics include PESQ-WB, STOI, log-likelihood ratio (LLR), DNSMOS OVRL, UTMOS, SpeechBERT, Levenshtein phoneme similarity (LPS), speaker similarity, and differential word error rate (dWER). Specifically, dWER is computed from transcriptions generated by Whisper large-v3; SpeechBERT is computed using the hidden representations from the 14th layer of WavLM-large; phoneme sequences for LPS are extracted using \texttt{wav2vec2-lv-60-espeak-cv-ft}; and speaker similarity is computed using \texttt{wavlm-base-plus-sv}.

\subsection{Speech Denoising and Joint Denoising and Dereverberation}
\nopagebreak[4]

\begin{appendixtable}
  \caption{Speech denoising performance on the Libri2Mix \textit{mix\_single} test set. The best results are shown in \textbf{bold}, and the second-best results are \underline{underlined}.}
  \label{tab:app_01}
  \begin{tabular}{@{}l*{9}{c}@{}}
    \toprule
    Model & \makecell{PESQ-WB\\$\uparrow$}
      & \makecell{STOI\\$\uparrow$} & \makecell{LLR\\$\downarrow$}
      & \makecell{DNSMOS\\OVRL $\uparrow$} & \makecell{UTMOS\\$\uparrow$}
      & \makecell{SpeechBERT\\$\uparrow$} & \makecell{LPS\\$\uparrow$}
      & \makecell{Spk. Sim.\\$\uparrow$} & \makecell{dWER\\$\downarrow$} \\
    \midrule
    AuK & 1.7871 & 81.64\% & 1.5325 & 3.2789 & 3.9927 & \underline{0.9125} & \textbf{0.9392} & \textbf{0.9665} & \textbf{6.12\%} \\
    Auphonic & 1.5554 & 75.04\% & 1.8720 & 3.3079 & 4.0885 & 0.8713 & 0.8869 & 0.9565 & 10.38\% \\
    UniSE & 1.6830 & 87.47\% & \textbf{0.6901} & \underline{3.3092} & \underline{4.1609} & 0.8842 & 0.8686 & 0.9462 & 15.41\% \\
    SAM Audio base & 1.8343 & \underline{89.64\%} & 0.8316 & 3.2056 & 3.4845 & 0.8669 & 0.8597 & 0.9573 & 14.15\% \\
    SAM Audio large & \underline{1.9189} & 88.16\% & 0.8641 & 3.1830 & 3.6586 & 0.8701 & 0.8487 & 0.9500 & 15.35\% \\
    \textbf{UniFE} & \textbf{2.1051} & \textbf{90.28\%} & \underline{0.7037} & \textbf{3.3556} & \textbf{4.2065} & \textbf{0.9178} & \underline{0.9372} & \underline{0.9655} & \underline{7.18\%} \\
    \bottomrule
  \end{tabular}
\end{appendixtable}

\begin{appendixtable}
  \caption{Speech denoising performance on the WHAM! \textit{mix\_single} test set. The best results are shown in \textbf{bold}, and the second-best results are \underline{underlined}.}
  \label{tab:app_02}
  \begin{tabular}{@{}l*{9}{c}@{}}
    \toprule
    Model & \makecell{PESQ-WB\\$\uparrow$}
      & \makecell{STOI\\$\uparrow$} & \makecell{LLR\\$\downarrow$}
      & \makecell{DNSMOS\\OVRL $\uparrow$} & \makecell{UTMOS\\$\uparrow$}
      & \makecell{SpeechBERT\\$\uparrow$} & \makecell{LPS\\$\uparrow$}
      & \makecell{Spk. Sim.\\$\uparrow$} & \makecell{dWER\\$\downarrow$} \\
    \midrule
    AuK & 1.7386 & \textbf{93.14\%} & 2.5479 & 3.1720 & 3.9411 & \textbf{0.9201} & \textbf{0.9389} & \textbf{0.9708} & \textbf{6.40\%} \\
    Auphonic & 1.5733 & 75.93\% & 1.9041 & 3.2367 & 4.0265 & 0.8739 & 0.8700 & 0.9580 & 10.77\% \\
    UniSE & 1.6406 & 88.71\% & \underline{0.7800} & \textbf{3.3804} & \underline{4.2211} & 0.8623 & 0.8363 & 0.9503 & 21.84\% \\
    SAM Audio base & 1.6981 & 90.18\% & 1.0814 & 3.1572 & 3.5554 & 0.8628 & 0.8325 & 0.9556 & 18.15\% \\
    SAM Audio large & \underline{1.7788} & 88.72\% & 1.1719 & 3.1117 & 3.7376 & 0.8647 & 0.8211 & 0.9473 & 19.82\% \\
    \textbf{UniFE} & \textbf{2.0468} & \underline{91.49\%} & \textbf{0.7080} & \underline{3.2968} & \textbf{4.2645} & \underline{0.9171} & \underline{0.9268} & \underline{0.9674} & \underline{9.97\%} \\
    \bottomrule
  \end{tabular}
\end{appendixtable}

\begin{appendixtable}
  \caption{Joint speech denoising and dereverberation performance on the WHAMR! \textit{mix\_single\_reverb} test set. The best results are shown in \textbf{bold}, and the second-best results are \underline{underlined}.}
  \label{tab:app_03}
  \begin{tabular}{@{}l*{9}{c}@{}}
    \toprule
    Model & \makecell{PESQ-WB\\$\uparrow$}
      & \makecell{STOI\\$\uparrow$} & \makecell{LLR\\$\downarrow$}
      & \makecell{DNSMOS\\OVRL $\uparrow$} & \makecell{UTMOS\\$\uparrow$}
      & \makecell{SpeechBERT\\$\uparrow$} & \makecell{LPS\\$\uparrow$}
      & \makecell{Spk. Sim.\\$\uparrow$} & \makecell{dWER\\$\downarrow$} \\
    \midrule
    AuK & 1.4610 & 85.02\% & 2.6667 & 2.7909 & 2.9505 & \underline{0.8784} & \underline{0.9125} & \underline{0.9579} & \textbf{7.53\%} \\
    Auphonic & 1.5160 & 75.61\% & 1.9504 & 3.2336 & 3.9244 & 0.8587 & 0.8372 & 0.9512 & 14.49\% \\
    UniSE & \underline{1.5697} & \underline{87.82\%} & \underline{0.8172} & \textbf{3.3733} & \underline{4.1918} & 0.8503 & 0.8095 & 0.9460 & 27.96\% \\
    SAM Audio base & 1.4560 & 82.63\% & 1.2219 & 2.7569 & 2.4842 & 0.8282 & 0.8061 & 0.9409 & 17.91\% \\
    SAM Audio large & 1.4889 & 81.05\% & 1.3392 & 2.7622 & 2.6845 & 0.8288 & 0.7919 & 0.9313 & 21.38\% \\
    \textbf{UniFE} & \textbf{1.9956} & \textbf{91.08\%} & \textbf{0.6793} & \underline{3.2983} & \textbf{4.2450} & \textbf{0.9113} & \textbf{0.9184} & \textbf{0.9628} & \underline{10.09\%} \\
    \bottomrule
  \end{tabular}
\end{appendixtable}

\begin{appendixtable}
  \caption{Speech denoising performance on the EARS-WHAM v2 test set. The best results are shown in \textbf{bold}, and the second-best results are \underline{underlined}.}
  \label{tab:app_04}
  \begin{tabular}{@{}l*{9}{c}@{}}
    \toprule
    Model & \makecell{PESQ-WB\\$\uparrow$}
      & \makecell{STOI\\$\uparrow$} & \makecell{LLR\\$\downarrow$}
      & \makecell{DNSMOS\\OVRL $\uparrow$} & \makecell{UTMOS\\$\uparrow$}
      & \makecell{SpeechBERT\\$\uparrow$} & \makecell{LPS\\$\uparrow$}
      & \makecell{Spk. Sim.\\$\uparrow$} & \makecell{dWER\\$\downarrow$} \\
    \midrule
    AuK & \textbf{2.1407} & \textbf{90.21\%} & 1.5758 & 3.1442 & 3.1998 & \textbf{0.9011} & \textbf{0.8863} & \textbf{0.9839} & \textbf{5.48\%} \\
    Auphonic & 1.8853 & 74.38\% & 1.4485 & 3.1863 & 3.2868 & 0.8670 & 0.8224 & \underline{0.9756} & \underline{8.41\%} \\
    UniSE & 1.6419 & 83.84\% & \underline{0.9248} & \textbf{3.3237} & \underline{3.4580} & 0.8323 & 0.7671 & 0.9594 & 21.11\% \\
    SAM Audio base & 1.9052 & 87.29\% & 0.9913 & 3.1244 & 2.8965 & 0.8540 & 0.7776 & 0.9686 & 14.92\% \\
    SAM Audio large & 2.0419 & \underline{88.08\%} & 0.9461 & 3.1558 & 3.1556 & 0.8701 & 0.8023 & 0.9728 & 13.93\% \\
    \textbf{UniFE} & \underline{2.0648} & 85.88\% & \textbf{0.8844} & \underline{3.2468} & \textbf{3.6198} & \underline{0.8840} & \underline{0.8418} & 0.9717 & 13.69\% \\
    \bottomrule
  \end{tabular}
\end{appendixtable}

\begin{appendixtable}
  \caption{Speech denoising performance on the \textit{denoising} subset of the MultiDistortion dataset. The best results are shown in \textbf{bold}, and the second-best results are \underline{underlined}.}
  \label{tab:app_05}
  \begin{tabular}{@{}l*{9}{c}@{}}
    \toprule
    Model & \makecell{PESQ-WB\\$\uparrow$}
      & \makecell{STOI\\$\uparrow$} & \makecell{LLR\\$\downarrow$}
      & \makecell{DNSMOS\\OVRL $\uparrow$} & \makecell{UTMOS\\$\uparrow$}
      & \makecell{SpeechBERT\\$\uparrow$} & \makecell{LPS\\$\uparrow$}
      & \makecell{Spk. Sim.\\$\uparrow$} & \makecell{dWER\\$\downarrow$} \\
    \midrule
    AuK & \underline{2.3085} & \textbf{93.10\%} & 1.8778 & 3.3398 & 3.5678 & \underline{0.9084} & \textbf{0.9047} & \textbf{0.9809} & \textbf{6.05\%} \\
    Auphonic & 1.8375 & 76.95\% & 1.5821 & 3.4225 & 3.8165 & 0.8651 & 0.8381 & 0.9703 & 17.87\% \\
    UniSE & 1.8458 & 88.53\% & \underline{0.7883} & \underline{3.4591} & \underline{3.8918} & 0.8639 & 0.8353 & 0.9611 & 18.77\% \\
    SAM Audio base & 2.0963 & 87.47\% & 1.0099 & 3.2564 & 3.3046 & 0.8400 & 0.7753 & 0.9409 & 27.56\% \\
    SAM Audio large & 2.2317 & 89.30\% & 0.9218 & 3.3294 & 3.5378 & 0.8607 & 0.8052 & 0.9593 & 32.58\% \\
    \textbf{UniFE} & \textbf{2.3456} & \underline{90.73\%} & \textbf{0.6107} & \textbf{3.5010} & \textbf{4.0838} & \textbf{0.9107} & \underline{0.9007} & \underline{0.9773} & \underline{8.60\%} \\
    \bottomrule
  \end{tabular}
\end{appendixtable}

\subsection{Target Speaker Extraction}
\nopagebreak[4]

\begin{appendixtable}
  \caption{TSE performance on the Libri2Mix \textit{mix\_clean} test set. The best results are shown in \textbf{bold}, and the second-best results are \underline{underlined}.}
  \label{tab:app_06}
  \begin{tabular}{@{}l*{9}{c}@{}}
    \toprule
    Model & \makecell{PESQ-WB\\$\uparrow$}
      & \makecell{STOI\\$\uparrow$} & \makecell{LLR\\$\downarrow$}
      & \makecell{DNSMOS\\OVRL $\uparrow$} & \makecell{UTMOS\\$\uparrow$}
      & \makecell{SpeechBERT\\$\uparrow$} & \makecell{LPS\\$\uparrow$}
      & \makecell{Spk. Sim.\\$\uparrow$} & \makecell{dWER\\$\downarrow$} \\
    \midrule
    LauraTSE & \textbf{2.4355} & \textbf{91.67\%} & \textbf{0.3987} & \underline{3.3474} & 4.1345 & \underline{0.9183} & \underline{0.9287} & \textbf{0.9743} & \textbf{5.87\%} \\
    UniSE & 1.6656 & 84.18\% & 0.6937 & \textbf{4.0702} & \underline{4.1605} & 0.8818 & 0.8137 & 0.9467 & 19.59\% \\
    \textbf{UniFE} & \underline{2.3630} & \underline{90.44\%} & \underline{0.5632} & 3.3409 & \textbf{4.1816} & \textbf{0.9281} & \textbf{0.9335} & \underline{0.9704} & \underline{6.99\%} \\
    \bottomrule
  \end{tabular}
\end{appendixtable}

\begin{appendixtable}
  \caption{TSE performance in noisy conditions on the Libri2Mix \textit{mix\_both} test set. The best results are shown in \textbf{bold}, and the second-best results are \underline{underlined}.}
  \label{tab:app_07}
  \begin{tabular}{@{}l*{9}{c}@{}}
    \toprule
    Model & \makecell{PESQ-WB\\$\uparrow$}
      & \makecell{STOI\\$\uparrow$} & \makecell{LLR\\$\downarrow$}
      & \makecell{DNSMOS\\OVRL $\uparrow$} & \makecell{UTMOS\\$\uparrow$}
      & \makecell{SpeechBERT\\$\uparrow$} & \makecell{LPS\\$\uparrow$}
      & \makecell{Spk. Sim.\\$\uparrow$} & \makecell{dWER\\$\downarrow$} \\
    \midrule
    LauraTSE & \underline{1.4716} & 74.06\% & 1.0346 & 2.9076 & 3.0394 & 0.7765 & 0.6105 & 0.9236 & \underline{44.19\%} \\
    UniSE & 1.3772 & \underline{76.63\%} & \underline{0.9814} & \underline{3.3430} & \underline{4.1156} & \underline{0.8152} & \underline{0.6196} & \underline{0.9250} & 48.84\% \\
    \textbf{UniFE} & \textbf{1.8800} & \textbf{86.78\%} & \textbf{0.7429} & \textbf{3.3498} & \textbf{4.1797} & \textbf{0.8903} & \textbf{0.8510} & \textbf{0.9606} & \textbf{18.79\%} \\
    \bottomrule
  \end{tabular}
\end{appendixtable}

\begin{appendixtable}
  \caption{TSE performance on the WHAM! \textit{mix\_clean} test set. The best results are shown in \textbf{bold}, and the second-best results are \underline{underlined}.}
  \label{tab:app_08}
  \begin{tabular}{@{}l*{9}{c}@{}}
    \toprule
    Model & \makecell{PESQ-WB\\$\uparrow$}
      & \makecell{STOI\\$\uparrow$} & \makecell{LLR\\$\downarrow$}
      & \makecell{DNSMOS\\OVRL $\uparrow$} & \makecell{UTMOS\\$\uparrow$}
      & \makecell{SpeechBERT\\$\uparrow$} & \makecell{LPS\\$\uparrow$}
      & \makecell{Spk. Sim.\\$\uparrow$} & \makecell{dWER\\$\downarrow$} \\
    \midrule
    LauraTSE & \textbf{2.7871} & \underline{91.52\%} & \textbf{0.3690} & \underline{3.3197} & 4.1657 & \underline{0.9319} & \underline{0.9503} & \textbf{0.9819} & \textbf{3.72\%} \\
    UniSE & 1.8890 & 88.19\% & 0.6608 & \textbf{3.3539} & \underline{4.1760} & 0.8913 & 0.8426 & 0.9601 & 16.64\% \\
    \textbf{UniFE} & \underline{2.5997} & \textbf{93.22\%} & \underline{0.4843} & 3.2478 & \textbf{4.2178} & \textbf{0.9395} & \textbf{0.9523} & \underline{0.9775} & \underline{4.19\%} \\
    \bottomrule
  \end{tabular}
\end{appendixtable}

\begin{appendixtable}
  \caption{TSE performance in noisy conditions on the WHAM! \textit{mix\_both} test set. The best results are shown in \textbf{bold}, and the second-best results are \underline{underlined}.}
  \label{tab:app_09}
  \begin{tabular}{@{}l*{9}{c}@{}}
    \toprule
    Model & \makecell{PESQ-WB\\$\uparrow$}
      & \makecell{STOI\\$\uparrow$} & \makecell{LLR\\$\downarrow$}
      & \makecell{DNSMOS\\OVRL $\uparrow$} & \makecell{UTMOS\\$\uparrow$}
      & \makecell{SpeechBERT\\$\uparrow$} & \makecell{LPS\\$\uparrow$}
      & \makecell{Spk. Sim.\\$\uparrow$} & \makecell{dWER\\$\downarrow$} \\
    \midrule
    LauraTSE & 1.3906 & 60.68\% & 1.3392 & 2.8776 & 2.8310 & 0.7358 & 0.4748 & 0.9003 & 66.73\% \\
    UniSE & \underline{1.3965} & \underline{80.75\%} & \underline{1.0042} & \textbf{3.3749} & \underline{4.1108} & \underline{0.8128} & \underline{0.6105} & \underline{0.9378} & \underline{58.57\%} \\
    \textbf{UniFE} & \textbf{1.8936} & \textbf{89.45\%} & \textbf{0.6608} & \underline{3.2587} & \textbf{4.2065} & \textbf{0.8975} & \textbf{0.8603} & \textbf{0.9666} & \textbf{19.96\%} \\
    \bottomrule
  \end{tabular}
\end{appendixtable}

\begin{appendixtable}
  \caption{TSE performance in reverberant conditions on the WHAMR! \textit{mix\_clean\_reverb} test set. The best results are shown in \textbf{bold}, and the second-best results are \underline{underlined}.}
  \label{tab:app_10}
  \begin{tabular}{@{}l*{9}{c}@{}}
    \toprule
    Model & \makecell{PESQ-WB\\$\uparrow$}
      & \makecell{STOI\\$\uparrow$} & \makecell{LLR\\$\downarrow$}
      & \makecell{DNSMOS\\OVRL $\uparrow$} & \makecell{UTMOS\\$\uparrow$}
      & \makecell{SpeechBERT\\$\uparrow$} & \makecell{LPS\\$\uparrow$}
      & \makecell{Spk. Sim.\\$\uparrow$} & \makecell{dWER\\$\downarrow$} \\
    \midrule
    LauraTSE & \underline{1.8279} & 78.75\% & \underline{0.8101} & 3.0918 & 3.5029 & 0.8503 & \underline{0.8042} & 0.9484 & \underline{19.98\%} \\
    UniSE & 1.6639 & \underline{83.97\%} & 0.8770 & \textbf{3.3697} & \underline{4.1443} & \underline{0.8528} & 0.7348 & \underline{0.9495} & 33.12\% \\
    \textbf{UniFE} & \textbf{2.4060} & \textbf{92.12\%} & \textbf{0.5196} & \underline{3.2382} & \textbf{4.2032} & \textbf{0.9276} & \textbf{0.9308} & \textbf{0.9724} & \textbf{7.54\%} \\
    \bottomrule
  \end{tabular}
\end{appendixtable}

\begin{appendixtable}
  \caption{TSE performance in noisy and reverberant conditions on the WHAMR! \textit{mix\_both\_reverb} test set. The best results are shown in \textbf{bold}, and the second-best results are \underline{underlined}.}
  \label{tab:app_11}
  \begin{tabular}{@{}l*{9}{c}@{}}
    \toprule
    Model & \makecell{PESQ-WB\\$\uparrow$}
      & \makecell{STOI\\$\uparrow$} & \makecell{LLR\\$\downarrow$}
      & \makecell{DNSMOS\\OVRL $\uparrow$} & \makecell{UTMOS\\$\uparrow$}
      & \makecell{SpeechBERT\\$\uparrow$} & \makecell{LPS\\$\uparrow$}
      & \makecell{Spk. Sim.\\$\uparrow$} & \makecell{dWER\\$\downarrow$} \\
    \midrule
    LauraTSE & \underline{1.3386} & 45.48\% & 1.6502 & 2.2440 & 1.9971 & 0.6715 & 0.3627 & 0.8455 & 77.24\% \\
    UniSE & 1.3288 & \underline{77.77\%} & \underline{1.1535} & \textbf{3.3854} & \underline{4.0787} & \underline{0.7888} & \underline{0.5303} & \underline{0.9330} & \underline{72.53\%} \\
    \textbf{UniFE} & \textbf{1.7808} & \textbf{87.98\%} & \textbf{0.7120} & \underline{3.2760} & \textbf{4.2075} & \textbf{0.8827} & \textbf{0.8219} & \textbf{0.9623} & \textbf{26.92\%} \\
    \bottomrule
  \end{tabular}
\end{appendixtable}

\subsection{Restoration of Individual Distortions}
\nopagebreak[4]

\begin{appendixtable}
  \caption{Speech dereverberation performance on the EARS-Reverb v2 test set. The best results are shown in \textbf{bold}, and the second-best results are \underline{underlined}.}
  \label{tab:app_12}
  \begin{tabular}{@{}l*{9}{c}@{}}
    \toprule
    Model & \makecell{PESQ-WB\\$\uparrow$}
      & \makecell{STOI\\$\uparrow$} & \makecell{LLR\\$\downarrow$}
      & \makecell{DNSMOS\\OVRL $\uparrow$} & \makecell{UTMOS\\$\uparrow$}
      & \makecell{SpeechBERT\\$\uparrow$} & \makecell{LPS\\$\uparrow$}
      & \makecell{Spk. Sim.\\$\uparrow$} & \makecell{dWER\\$\downarrow$} \\
    \midrule
    AuK & \textbf{2.5468} & 77.13\% & 0.9373 & \underline{3.2549} & \underline{3.3134} & \textbf{0.9260} & \textbf{0.9315} & \textbf{0.9872} & \textbf{1.30\%} \\
    Auphonic & 1.9351 & 74.26\% & 1.2119 & 3.2212 & \textbf{3.3407} & 0.8909 & 0.8772 & \underline{0.9776} & \underline{2.92\%} \\
    UniSE & 1.5892 & \underline{84.29\%} & \underline{0.7832} & \textbf{3.3294} & 3.2823 & 0.8621 & 0.8525 & 0.9621 & 7.59\% \\
    SAM Audio base & 1.2491 & 54.18\% & 2.0268 & 1.9900 & 2.1646 & 0.6166 & 0.3654 & 0.7794 & 57.29\% \\
    SAM Audio large & 1.2500 & 43.80\% & 2.3786 & 1.9775 & 2.7881 & 0.5298 & 0.1455 & 0.6491 & 86.21\% \\
    \textbf{UniFE} & \underline{2.0743} & \textbf{86.86\%} & \textbf{0.7804} & 3.0565 & 3.1345 & \underline{0.9026} & \underline{0.9041} & 0.9754 & 2.99\% \\
    \bottomrule
  \end{tabular}
\end{appendixtable}

\begin{appendixtable}
  \caption{Speech dereverberation performance on the WHAMR! \textit{s1\_reverb} test set. The best results are shown in \textbf{bold}, and the second-best results are \underline{underlined}.}
  \label{tab:app_13}
  \begin{tabular}{@{}l*{9}{c}@{}}
    \toprule
    Model & \makecell{PESQ-WB\\$\uparrow$}
      & \makecell{STOI\\$\uparrow$} & \makecell{LLR\\$\downarrow$}
      & \makecell{DNSMOS\\OVRL $\uparrow$} & \makecell{UTMOS\\$\uparrow$}
      & \makecell{SpeechBERT\\$\uparrow$} & \makecell{LPS\\$\uparrow$}
      & \makecell{Spk. Sim.\\$\uparrow$} & \makecell{dWER\\$\downarrow$} \\
    \midrule
    AuK & \underline{2.7395} & 54.83\% & 1.6067 & 3.2891 & 4.0697 & \underline{0.9409} & \underline{0.9625} & \textbf{0.9828} & \underline{2.75\%} \\
    Auphonic & 1.9621 & 77.54\% & 1.6868 & \underline{3.3185} & \textbf{4.3078} & 0.9267 & 0.9558 & 0.9735 & 2.81\% \\
    UniSE & 2.2176 & \underline{92.58\%} & \underline{0.4524} & \textbf{3.3421} & \underline{4.2540} & 0.9160 & 0.9523 & 0.9684 & 4.26\% \\
    SAM Audio base & 2.0889 & 89.69\% & 0.6793 & 2.7174 & 3.0099 & 0.9294 & 0.9580 & 0.9743 & \textbf{2.53\%} \\
    SAM Audio large & 2.0944 & 87.82\% & 0.7330 & 2.7212 & 3.0260 & 0.9180 & 0.9249 & 0.9602 & 5.04\% \\
    \textbf{UniFE} & \textbf{2.8294} & \textbf{94.11\%} & \textbf{0.3994} & 3.2396 & 4.2165 & \textbf{0.9523} & \textbf{0.9662} & \underline{0.9785} & 2.90\% \\
    \bottomrule
  \end{tabular}
\end{appendixtable}

\begin{appendixtable}
  \caption{Speech dereverberation performance on the \textit{dereverb} subset of the MultiDistortion dataset. The best results are shown in \textbf{bold}, and the second-best results are \underline{underlined}.}
  \label{tab:app_14}
  \begin{tabular}{@{}l*{9}{c}@{}}
    \toprule
    Model & \makecell{PESQ-WB\\$\uparrow$}
      & \makecell{STOI\\$\uparrow$} & \makecell{LLR\\$\downarrow$}
      & \makecell{DNSMOS\\OVRL $\uparrow$} & \makecell{UTMOS\\$\uparrow$}
      & \makecell{SpeechBERT\\$\uparrow$} & \makecell{LPS\\$\uparrow$}
      & \makecell{Spk. Sim.\\$\uparrow$} & \makecell{dWER\\$\downarrow$} \\
    \midrule
    AuK & \textbf{2.4711} & 82.09\% & 1.0390 & 3.4366 & 3.7963 & \underline{0.9224} & \textbf{0.9377} & \underline{0.9820} & \textbf{1.49\%} \\
    Auphonic & 1.8716 & 78.15\% & 1.5748 & \textbf{3.4692} & \underline{3.9563} & 0.8924 & 0.9082 & 0.9761 & \underline{3.11\%} \\
    UniSE & 1.7625 & \underline{87.90\%} & \underline{0.8060} & 3.4472 & 3.7851 & 0.8932 & 0.9044 & 0.9699 & 8.70\% \\
    SAM Audio base & 1.2997 & 55.82\% & 1.9378 & 2.1828 & 2.2380 & 0.6496 & 0.4267 & 0.7285 & 58.89\% \\
    SAM Audio large & 1.3537 & 53.03\% & 1.8172 & 2.3160 & 2.4923 & 0.6401 & 0.3803 & 0.7056 & 62.93\% \\
    \textbf{UniFE} & \underline{2.4301} & \textbf{90.69\%} & \textbf{0.5158} & \underline{3.4596} & \textbf{3.9571} & \textbf{0.9305} & \underline{0.9300} & \textbf{0.9824} & 3.19\% \\
    \bottomrule
  \end{tabular}
\end{appendixtable}

\begin{appendixtable}
  \caption{Packet loss concealment performance on the PLC-Challenge 22 \textit{blind} test set. The best results are shown in \textbf{bold}, and the second-best results are \underline{underlined}.}
  \label{tab:app_15}
  \begin{tabular}{@{}l*{9}{c}@{}}
    \toprule
    Model & \makecell{PESQ-WB\\$\uparrow$}
      & \makecell{STOI\\$\uparrow$} & \makecell{LLR\\$\downarrow$}
      & \makecell{DNSMOS\\OVRL $\uparrow$} & \makecell{UTMOS\\$\uparrow$}
      & \makecell{SpeechBERT\\$\uparrow$} & \makecell{LPS\\$\uparrow$}
      & \makecell{Spk. Sim.\\$\uparrow$} & \makecell{dWER\\$\downarrow$} \\
    \midrule
    AuK & 1.9757 & 81.11\% & \textbf{0.4573} & 2.7580 & 2.3569 & \underline{0.8848} & \underline{0.8162} & 0.9616 & \textbf{9.26\%} \\
    Auphonic & 1.4385 & 64.62\% & 1.7759 & 3.1455 & 3.0233 & 0.8439 & 0.7837 & 0.9526 & 15.32\% \\
    UniSE & 1.7341 & \underline{82.76\%} & 0.7767 & \underline{3.3395} & \underline{3.6620} & 0.8798 & 0.8077 & \underline{0.9645} & 21.23\% \\
    SAM Audio base & 2.0391 & 81.49\% & \underline{0.6461} & 2.8003 & 2.5556 & 0.8847 & 0.8028 & 0.9543 & 14.23\% \\
    SAM Audio large & \underline{2.0633} & 80.86\% & 0.7141 & 2.8191 & 2.6274 & 0.8739 & 0.7922 & 0.9426 & 16.96\% \\
    \textbf{UniFE} & \textbf{2.2222} & \textbf{86.22\%} & 0.7236 & \textbf{3.3693} & \textbf{3.7308} & \textbf{0.9130} & \textbf{0.8660} & \textbf{0.9798} & \underline{12.98\%} \\
    \bottomrule
  \end{tabular}
\end{appendixtable}

\begin{appendixtable}
  \caption{Packet loss concealment performance on the PLC-Challenge 24 \textit{val\_set} v2. The best results are shown in \textbf{bold}, and the second-best results are \underline{underlined}.}
  \label{tab:app_16}
  \begin{tabular}{@{}l*{9}{c}@{}}
    \toprule
    Model & \makecell{PESQ-WB\\$\uparrow$}
      & \makecell{STOI\\$\uparrow$} & \makecell{LLR\\$\downarrow$}
      & \makecell{DNSMOS\\OVRL $\uparrow$} & \makecell{UTMOS\\$\uparrow$}
      & \makecell{SpeechBERT\\$\uparrow$} & \makecell{LPS\\$\uparrow$}
      & \makecell{Spk. Sim.\\$\uparrow$} & \makecell{dWER\\$\downarrow$} \\
    \midrule
    AuK & 1.7490 & 75.46\% & \textbf{0.4490} & 2.8466 & 2.3271 & 0.8584 & 0.7713 & 0.9687 & \underline{19.44\%} \\
    Auphonic & 1.3327 & 61.10\% & 1.6345 & 3.1897 & 2.9578 & 0.8269 & 0.7513 & 0.9595 & 23.90\% \\
    UniSE & 1.6093 & \underline{80.63\%} & 0.6571 & \underline{3.3669} & \underline{3.7676} & \underline{0.8728} & \underline{0.7952} & \underline{0.9738} & 31.17\% \\
    SAM Audio base & 1.7830 & 76.29\% & 0.6916 & 2.9434 & 2.4978 & 0.8588 & 0.7647 & 0.9589 & 21.80\% \\
    SAM Audio large & \underline{1.8211} & 74.89\% & 0.8091 & 2.9603 & 2.5981 & 0.8363 & 0.7278 & 0.9324 & 28.69\% \\
    \textbf{UniFE} & \textbf{2.0112} & \textbf{83.67\%} & \underline{0.6557} & \textbf{3.4067} & \textbf{3.8074} & \textbf{0.9050} & \textbf{0.8432} & \textbf{0.9852} & \textbf{19.09\%} \\
    \bottomrule
  \end{tabular}
\end{appendixtable}

\begin{appendixtable}
  \caption{Packet loss concealment performance on the \textit{packet\_loss} subset of the MultiDistortion dataset. The best results are shown in \textbf{bold}, and the second-best results are \underline{underlined}.}
  \label{tab:app_17}
  \begin{tabular}{@{}l*{9}{c}@{}}
    \toprule
    Model & \makecell{PESQ-WB\\$\uparrow$}
      & \makecell{STOI\\$\uparrow$} & \makecell{LLR\\$\downarrow$}
      & \makecell{DNSMOS\\OVRL $\uparrow$} & \makecell{UTMOS\\$\uparrow$}
      & \makecell{SpeechBERT\\$\uparrow$} & \makecell{LPS\\$\uparrow$}
      & \makecell{Spk. Sim.\\$\uparrow$} & \makecell{dWER\\$\downarrow$} \\
    \midrule
    AuK & 1.6874 & 82.44\% & \textbf{0.3314} & 3.2133 & 2.8505 & \underline{0.9077} & \underline{0.9013} & \textbf{0.9909} & \textbf{5.10\%} \\
    Auphonic & 1.4305 & 66.79\% & 1.3894 & 3.3616 & 3.5268 & 0.8740 & 0.8516 & 0.9813 & \underline{7.59\%} \\
    UniSE & \underline{1.7872} & \underline{88.14\%} & 0.5250 & \textbf{3.4978} & \underline{3.9817} & 0.9011 & 0.8828 & 0.9799 & 13.22\% \\
    SAM Audio base & 1.5074 & 67.42\% & 1.2556 & 2.8117 & 3.0851 & 0.7526 & 0.5921 & 0.8332 & 38.79\% \\
    SAM Audio large & 1.7398 & 81.63\% & 0.6146 & 3.2471 & 2.9758 & 0.9020 & 0.8706 & 0.9845 & 8.82\% \\
    \textbf{UniFE} & \textbf{2.2369} & \textbf{90.83\%} & \underline{0.3994} & \underline{3.4814} & \textbf{4.0519} & \textbf{0.9334} & \textbf{0.9125} & \underline{0.9876} & 9.13\% \\
    \bottomrule
  \end{tabular}
\end{appendixtable}

\begin{appendixtable}
  \caption{Interfering-speaker suppression performance on the \textit{interference} subset of the MultiDistortion dataset. The best results are shown in \textbf{bold}, and the second-best results are \underline{underlined}.}
  \label{tab:app_18}
  \begin{tabular}{@{}l*{9}{c}@{}}
    \toprule
    Model & \makecell{PESQ-WB\\$\uparrow$}
      & \makecell{STOI\\$\uparrow$} & \makecell{LLR\\$\downarrow$}
      & \makecell{DNSMOS\\OVRL $\uparrow$} & \makecell{UTMOS\\$\uparrow$}
      & \makecell{SpeechBERT\\$\uparrow$} & \makecell{LPS\\$\uparrow$}
      & \makecell{Spk. Sim.\\$\uparrow$} & \makecell{dWER\\$\downarrow$} \\
    \midrule
    AuK & 2.3957 & \underline{93.82\%} & 0.8333 & 3.1780 & 2.7490 & 0.8864 & 0.7572 & 0.9809 & 3.56\% \\
    Auphonic & 1.7317 & 77.34\% & 1.6044 & 3.3188 & 2.8846 & 0.8504 & 0.7292 & 0.9773 & \underline{2.31\%} \\
    UniSE & 2.3192 & 92.39\% & \underline{0.4969} & \textbf{3.4947} & \underline{3.9800} & \underline{0.9199} & \underline{0.9328} & \underline{0.9817} & 3.35\% \\
    SAM Audio base & \underline{2.7836} & 76.61\% & 1.0126 & 3.1496 & 3.2965 & 0.8479 & 0.7032 & 0.9012 & 26.66\% \\
    SAM Audio large & 2.5370 & 74.33\% & 1.1164 & 3.2002 & 3.1464 & 0.8489 & 0.6713 & 0.9138 & 32.16\% \\
    \textbf{UniFE} & \textbf{2.9281} & \textbf{94.07\%} & \textbf{0.3616} & \underline{3.4884} & \textbf{4.0515} & \textbf{0.9469} & \textbf{0.9429} & \textbf{0.9884} & \textbf{1.78\%} \\
    \bottomrule
  \end{tabular}
\end{appendixtable}

\begin{appendixtable}
  \caption{Speech bandwidth extension performance on the \textit{bandwidth\_limitation} subset of the MultiDistortion dataset. The best results are shown in \textbf{bold}, and the second-best results are \underline{underlined}.}
  \label{tab:app_19}
  \begin{tabular}{@{}l*{9}{c}@{}}
    \toprule
    Model & \makecell{PESQ-WB\\$\uparrow$}
      & \makecell{STOI\\$\uparrow$} & \makecell{LLR\\$\downarrow$}
      & \makecell{DNSMOS\\OVRL $\uparrow$} & \makecell{UTMOS\\$\uparrow$}
      & \makecell{SpeechBERT\\$\uparrow$} & \makecell{LPS\\$\uparrow$}
      & \makecell{Spk. Sim.\\$\uparrow$} & \makecell{dWER\\$\downarrow$} \\
    \midrule
    AuK & \underline{2.7999} & \underline{93.27\%} & 4.0388 & 3.2589 & 3.2202 & 0.9190 & \underline{0.9279} & 0.9512 & \textbf{2.79\%} \\
    Auphonic & 1.9766 & 77.00\% & 1.4840 & \underline{3.4777} & \underline{3.9803} & 0.9029 & 0.9171 & 0.9584 & \underline{3.24\%} \\
    UniSE & 2.1585 & 91.01\% & \underline{0.7353} & 3.4724 & 3.9030 & 0.9087 & 0.9171 & \underline{0.9625} & 5.82\% \\
    SAM Audio base & 2.7481 & 76.30\% & 2.7257 & 2.9341 & 3.2102 & 0.7703 & 0.6100 & 0.7993 & 35.48\% \\
    SAM Audio large & \textbf{3.5448} & \textbf{93.79\%} & 2.8185 & 3.4019 & 3.2035 & \underline{0.9194} & 0.8937 & 0.9397 & 3.88\% \\
    \textbf{UniFE} & 2.6761 & 92.92\% & \textbf{0.6409} & \textbf{3.4799} & \textbf{4.0537} & \textbf{0.9393} & \textbf{0.9371} & \textbf{0.9773} & 3.43\% \\
    \bottomrule
  \end{tabular}
\end{appendixtable}

\begin{appendixtable}
  \caption{Speech declipping performance on the \textit{clipping} subset of the MultiDistortion dataset. The best results are shown in \textbf{bold}, and the second-best results are \underline{underlined}.}
  \label{tab:app_20}
  \begin{tabular}{@{}l*{9}{c}@{}}
    \toprule
    Model & \makecell{PESQ-WB\\$\uparrow$}
      & \makecell{STOI\\$\uparrow$} & \makecell{LLR\\$\downarrow$}
      & \makecell{DNSMOS\\OVRL $\uparrow$} & \makecell{UTMOS\\$\uparrow$}
      & \makecell{SpeechBERT\\$\uparrow$} & \makecell{LPS\\$\uparrow$}
      & \makecell{Spk. Sim.\\$\uparrow$} & \makecell{dWER\\$\downarrow$} \\
    \midrule
    AuK & 1.9420 & 90.52\% & 0.6136 & 3.3735 & 2.7109 & \underline{0.9358} & \underline{0.9341} & 0.9707 & 4.01\% \\
    Auphonic & 2.1727 & 77.34\% & 1.3459 & 3.4390 & \textbf{4.0463} & 0.9154 & 0.9339 & \underline{0.9811} & \underline{2.15\%} \\
    UniSE & \underline{2.2011} & \underline{91.43\%} & \underline{0.4851} & \underline{3.4639} & 3.9361 & 0.9168 & 0.9271 & 0.9772 & 3.37\% \\
    SAM Audio base & 1.8000 & 78.64\% & 0.9475 & 3.1224 & 2.8381 & 0.8506 & 0.7396 & 0.8762 & 24.27\% \\
    SAM Audio large & 1.8674 & 80.56\% & 0.8530 & 3.2220 & 2.8361 & 0.8755 & 0.7729 & 0.9163 & 17.87\% \\
    \textbf{UniFE} & \textbf{2.9020} & \textbf{93.37\%} & \textbf{0.3725} & \textbf{3.4830} & \underline{4.0446} & \textbf{0.9476} & \textbf{0.9403} & \textbf{0.9869} & \textbf{2.07\%} \\
    \bottomrule
  \end{tabular}
\end{appendixtable}

\subsection{Restoration under Combined Distortions}
\nopagebreak[4]

\begin{appendixtable}
  \caption{Speech restoration performance on the \textit{combined} subset of the MultiDistortion dataset. The input speech in this subset is affected by additive noise, interfering speakers, reverberation, clipping, bandwidth limitation, and packet loss. The best results are shown in \textbf{bold}, and the second-best results are \underline{underlined}.}
  \label{tab:app_21}
  \begin{tabular}{@{}l*{9}{c}@{}}
    \toprule
    Model & \makecell{PESQ-WB\\$\uparrow$}
      & \makecell{STOI\\$\uparrow$} & \makecell{LLR\\$\downarrow$}
      & \makecell{DNSMOS\\OVRL $\uparrow$} & \makecell{UTMOS\\$\uparrow$}
      & \makecell{SpeechBERT\\$\uparrow$} & \makecell{LPS\\$\uparrow$}
      & \makecell{Spk. Sim.\\$\uparrow$} & \makecell{dWER\\$\downarrow$} \\
    \midrule
    AuK & 1.1259 & 60.35\% & 3.0574 & 3.1163 & 2.4190 & \underline{0.7378} & \underline{0.5333} & \underline{0.8994} & \textbf{53.54\%} \\
    Auphonic & 1.1382 & 48.97\% & 2.3700 & 3.0358 & 2.4980 & 0.6754 & 0.3177 & 0.8449 & 83.10\% \\
    UniSE & \underline{1.1533} & \underline{65.74\%} & \underline{1.9053} & \underline{3.3756} & \underline{3.1411} & 0.7060 & 0.3335 & 0.8539 & 92.34\% \\
    SAM Audio base & 1.1490 & 34.17\% & 4.4405 & 1.4523 & 1.6181 & 0.4754 & 0.0712 & 0.4718 & 96.99\% \\
    SAM Audio large & 1.1199 & 35.07\% & 4.0319 & 1.6145 & 2.1207 & 0.4693 & 0.0677 & 0.5040 & 93.05\% \\
    \textbf{UniFE} & \textbf{1.3795} & \textbf{75.86\%} & \textbf{1.6205} & \textbf{3.4448} & \textbf{3.8140} & \textbf{0.8028} & \textbf{0.6104} & \textbf{0.9299} & \underline{57.42\%} \\
    \bottomrule
  \end{tabular}
\end{appendixtable}

\end{document}